\documentclass[11pt]{article}
\usepackage{BSMvietnam,epsfig}
\usepackage{amsmath,amssymb,hyperref}

\def\be{\begin{equation}}
\def\ee{\end{equation}}
\def\bea{\begin{eqnarray}}
\def\eea{\end{eqnarray}}

\def\bm#1{\mbox{\boldmath$#1$\unboldmath}} 

\def \beq{\begin{equation}}
\def \eeq{\end{equation}}
\def \bea{\begin{eqnarray}}
\def \eea{\end{eqnarray}}

\def \TeV{\, {\rm TeV}}

\def \Op{Q}

\begin{document}
\vspace*{4cm}

\title{REVIEW ON NEW PHYSICS IN HEAVY FLAVORS:  $\bm B$-MESON SECTOR}

\author{ULRICH HAISCH}

\address{Rudolf Peierls Centre for Theoretical Physics, University of Oxford, OX1 3PN Oxford, United Kingdom}

\maketitle\abstracts{A brief review of  theoretical aspects of new-physics searches in  flavor physics is given. Special attention is thereby devoted to $B_{s}$ mixing and the interplay of high- and low-$p_T$ observations.}

\section{Introduction}

The combined $2.1 \, {\rm fb}^{-1}$ of 2011 and 2012 LHCb data represent both a great success and a big disappointment: a success because   analyses of the data sets led to  textbook measurements of $B$-meson observables such as $B_s$--$\bar B_s$ mixing\hspace{1mm}\cite{bmix} and $B \to K^\ast \mu^+ \mu^-$;\hspace{1mm}\cite{brare} a disappointment because the wealth of data agrees well with the standard model (SM) predictions in essentially all channels, casting doubt on some of the  $(2\hspace{0.25mm}$--$\hspace{0.25mm} 4) \hspace{0.25mm} \sigma$  anomalies reported by the $B$ factories and the Tevatron experiments. The success story of LHCb culminated recently in finding the first evidence for the decay $B_s \to \mu^+ \mu^-$.\hspace{1mm}\cite{Aaij:2012nna} While the observation of $B_s \to \mu^+ \mu^-$ with a branching fraction close to the SM expectation excludes the possibility of spectacular new-physics (NP) effects in the $B$-meson sector, deviations of ${\cal O} (50\%)$, {\it i.e.}, NP effects of  ``natural" size, are only started to being probed by LHCb as well as ATLAS and CMS. The measurements of $B \to K^\ast \mu^+ \mu^-$ and the evidence for $B_s \to \mu^+ \mu^-$  hence mark the beginning of the flavor precision era at the LHC. 

\section{NP in $\bm B_s$--$\bm{\bar B_s}$  mixing}

The phenomenon of $B_s$--$\bar B_s$ oscillations is encoded in the elements $M_{12}^s$ and $\Gamma_{12}^s$ of the hermitian mass  and decay rate  matrices. In the SM, $M_{12}^s$ is calculated from the dispersive part of electroweak box diagrams with ``off-shell" top quarks, while $\Gamma_{12}^s$ is due to the absorptive part related to ``on-shell" light up-type quarks. Sufficiently heavy  NP in the off-diagonal element $M_{12}^s$  ($\Gamma_{12}^s$) can be described via effective $\Delta B =2$ ($\Delta B = 1$) interactions. Schematically, one has 
\begin{equation} \label{eq1} 
\begin{split}
& \hspace{5mm}  (M_{12}^s)_{\rm NP} \propto  C_2^i \, \left [ \; \raisebox{-7.5mm}{\includegraphics[height=1.75cm]{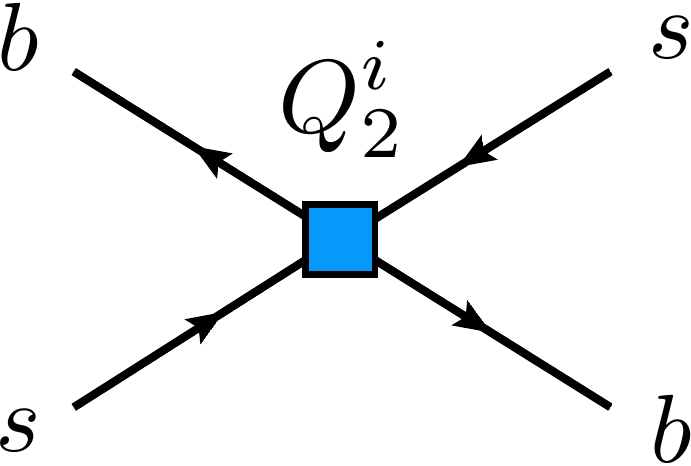}} \; \right ] \sim \frac{1}{\Lambda_{\rm NP}^2} \,,  \\[3mm] & \hspace{-5mm}  (\Gamma_{12}^s)_{\rm NP} \propto  C_1^i C_1^j \; {\rm Im} \left [ \; \raisebox{-10.5mm}{\includegraphics[height=2.2cm]{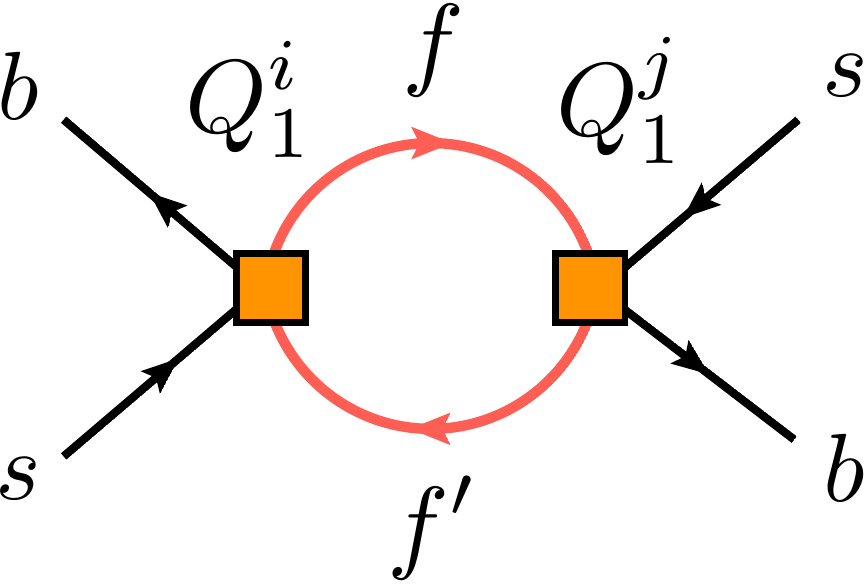}} \; \right ] \sim \frac{1}{(4\pi)^2} \frac{1}{\Lambda_{\rm NP}^4} \,, 
\end{split}
\end{equation}
where $\Lambda_{\rm NP}$ denotes the suppression scale of the higher-dimensional operators. 

The off-diagonal elements $M_{12}^s$ and $\Gamma_{12}^s$ can be related to the physical observables in the $B_s$-meson sector by introducing the following  model-independent parameterization:
\begin{equation} \label{eq2}
\begin{aligned} 
M_{12}^s = (M_{12}^s)_{\rm SM} + (M_{12}^s)_{\rm NP} & = (M_{12}^s)_{\rm SM} \, R_{M} \, e^{i \phi_{M}} \,, \\[1mm] \Gamma_{12}^s = (\Gamma_{12}^s)_{\rm SM} + (\Gamma_{12}^s)_{\rm NP} & = (\Gamma_{12}^s)_{\rm SM} \, R_{\Gamma} \, e^{i \phi_{\Gamma}} \,.
\end{aligned}
\end{equation}
To leading power in $|\Gamma_{12}^s|/|M_{12}^s|$ the mass difference $\Delta M_s$, the CP-violating phase $\phi_{J/\psi \phi}^s$, the decay-width difference $\Delta \Gamma_s$, and the flavor-specific CP asymmetry $a_{fs}^s$  are then given by 
\begin{gather} 
\Delta M_s = (\Delta M_s)_{\rm SM} \hspace{0.5mm} R_{M} \,, \qquad \phi_{J/\psi \phi}^s = (\phi_{J/\psi \phi}^s)_{\rm SM} + \phi_{M} \,, \nonumber \\[1mm]  \label{eq3} \\[-5mm] \Delta \Gamma_s \approx (\Delta \Gamma_s)_{\rm SM} \hspace{0.5mm} R_{\Gamma} \, \cos \left ( \phi_{M} - \phi_{\Gamma} \right ) \,, \quad a_{fs}^s \approx (a_{fs}^s)_{\rm SM}\, \frac{R_{\Gamma}}{R_{M}} \, \frac{\sin \left (\phi_{M} - \phi_{\Gamma} \right )}{ \phi_{\rm SM}^s} \,, \nonumber
\end{gather}
with $\phi_{\rm SM}^s = (0.22 \pm 0.06)^\circ$.\hspace{1mm}\cite{Lenz:2010gu} 

The $\Lambda_{\rm NP}$-scaling (\ref{eq1}) combined with (\ref{eq2}) and (\ref{eq3}) suggests that new particles are more likely to leave a visible imprint in $\Delta M_s$ and $\phi_{J/\psi \phi}^s$ than in $\Delta \Gamma_s$ and $a^s_{fs}$. Experimentally, however, the mass difference and the CP phase in $B_s$--$\bar B_s$ mixing are spot on the SM predictions,\hspace{1mm}\cite{bmix} while  $a^s_{fs}$, if extracted from the latest measurement\hspace{1mm}\cite{Abazov:2011yk}  of the like-sign dimuon charge asymmetry $A_{\rm SL}^b$, assuming the absence of NP in the $B_d$-meson system, shows an anomaly of almost $4\sigma$. In view of these results, it seems worthwhile to ask: how big can NP in $\Gamma_{12}^s$ be afterall? 

While any operator $(\bar s b) f$ with $f$ leading to a flavor-neutral final state of two or more fields and mass below the bottom-quark threshold can alter $\Gamma_{12}^s$, the possible final states are in practice limited, because essentially all $B_s \to f$  and $B_d \to X_s f$ decay modes involving light states in the final state are strongly constrained experimentally. An exception are $B$ decays to tau pairs.\hspace{1mm}\cite{Dighe:2007gt} The possibility of large  $b \to s \tau^+ \tau^-$ contributions to $\Gamma_{12}^s$, can be analyzed in a model-independent fashion\hspace{1mm}\cite{Bobeth:2011st} by adding the complete set of dimension-six operators ($A, B = L, R$)
\beq \label{eq4}
\begin{split}
  \Op_{S, AB} & = \left (\bar s \, P_A \, b \right ) \left (\bar \tau
    \, P_B \, \tau \right ) \,, \\
  \Op_{V, AB} & = \left (\bar s \, \gamma^\mu P_A \, b \right )
  \left (\bar \tau\, \gamma_\mu P_B \, \tau \right ) \,, \\
  \Op_{T, A} & = \left (\bar s \, \sigma^{\mu\nu} P_A \, b \right )
  \left (\bar \tau \, \sigma_{\mu \nu} P_A \,\tau \right ) \,,
\end{split}
\eeq
to the SM Lagrangian.  Here $P_{L,R} = (1 \mp \gamma_5)/2$ project onto left- and right-handed chiral fields and $\sigma^{\mu \nu} = i \left [\gamma^\mu, \gamma^\nu \right]/2$. 

The ten operators entering (\ref{eq4}) govern the purely leptonic $B_s \to \tau^+ \tau^-$ decay, the inclusive semileptonic $B \to X_s \tau^+\tau^-$ decay, and its exclusive counterpart $B^+ \to K^+ \tau^+\tau^-$, making these channels potentially powerful probes of $(\bar s b) (\bar \tau \tau)$ operators. In practice, these direct constraints however turn out to be rather loose at present. Explicitly, one obtains
\beq \label{eq:directBR}
\begin{split}
{\rm BR}(B_s \to \tau^+\tau^-) \, & < \, 3\% \,, \\
{\rm BR} (B \to X_s\tau^+\tau^-) \, & \lesssim \, 2.5 \% \,, \\
{\rm BR} (B^+ \to K^+ \tau^+\tau^-) \, & < \, 3.3 \cdot 10^{-3} \,.
\end{split}
\eeq
Here the first limit derives\hspace{1mm}\cite{Dighe:2010nj} from comparing the SM prediction $\tau_{B_s}/\tau_{B_d} - 1 \in [-0.4, 0.0] \%\;$\cite{Lenz:2012mb} with the corresponding experimental result $\tau_{B_s}/\tau_{B_d} - 1 = (0.4 \pm 1.9) \%$, while the second (crude) bound follows from estimating\hspace{1mm}\cite{Bobeth:2011st} the possible contamination of the exclusive and inclusive semileptonic decay samples by $B \to X_s \tau^+ \tau^-$ events. The final number corresponds to the 90\% confidence level (CL) upper limit on the branching ratio of $B^+ \to K^+ \tau^+ \tau^-$ as measured by BaBar.\hspace{1mm}\cite{Flood:2010zz}

Further constraints on the Wilson coefficients of the $(\bar s b)(\bar \tau \tau)$ operators arise indirectly from the experimentally available information on the $b \to s \gamma$, $b \to s \ell^+ \ell^-$ ($\ell = e, \mu$), and $b \to s \gamma \gamma$ transitions, because some of the effective operators introduced in (\ref{eq4}) mix into the electromagnetic dipole operators $Q_{7,A}$ and the vector-like semileptonic operators $Q_{9,A}$. An explicit calculation\hspace{1mm}\cite{Bobeth:2011st} shows that the operators $\Op_{S,AB}$ mix neither into $\Op_{7,A}$ nor $\Op_{9,A}$, while $\Op_{V,AB}$ ($\Op_{T,A}$) mixes only into $Q_{9,A}$ ($\Op_{7,A}$). As a result of the particular mixing pattern, the stringent constraints from the radiative decay $B \to X_s \gamma$ rule out large contributions to $\Gamma_{12}^s$ only if they arise from the tensor operators $\Op_{T,A}$.  Similarly, the rare decays $B \to X_s \ell^+ \ell^-$ and $B \to K^{(\ast)} \ell^+ \ell^-$ primarily limit contributions stemming from the vector operators $\Op_{V,AB}$. In contrast to $B \to X_s \gamma$, all $(\bar s b) (\bar \tau \tau)$ operators contribute to the double-radiative $B_s \to \gamma \gamma$ decay at the one-loop level. A detailed study\hspace{1mm}\cite{Bobeth:2011st} shows however  that the limits following from $b \to s \gamma \gamma$ are in practice not competitive with the bounds obtained  from the other tree- and loop-level mediated $B_{s,d}$-meson decays. 

The off-diagonal element $\Gamma_{12}^s$ is related via the optical theorem to the absorptive part of the forward-scattering amplitude which converts a $\bar B_s$ into a $B_s$ meson. Working to leading order in the strong coupling constant and $\Lambda_{\rm QCD}/m_b$, the contributions from the operators (\ref{eq4}) to $\Gamma_{12}^s$ is found by computing the matrix elements of the $(Q_i, Q_j)$ double insertions between quark states. Such a calculation\hspace{1mm}\cite{Bobeth:2011st} leads to the following  90\%~CL bounds 
\beq \label{eq:RGammaSVTbounds} 
(R_\Gamma)_{S,AB} < 1.15 \,,  \qquad (R_\Gamma)_{V,AB} < 1.35 \,, \qquad
(R_\Gamma)_{T,L} < 1.02 \,, \qquad (R_\Gamma)_{T,R} < 1.04 \,.
\eeq
These numbers imply that $(\bar s b) (\bar \tau \tau)$ operators of scalar (vector) type can lead to enhancements of $|\Gamma_{12}^s|$ over its SM value by 15\% (35\%) without violating any existing constraint. In contrast, contributions from tensor operators can alter $|\Gamma_{12}^s|$ by at most $4\%$.  Since an explanation of  the experimentally observed large negative values of $a_{fs}^s$ (or equivalent $A_{\rm SL}^b$) calls for NP in $\Gamma_{12}^s$ that changes the SM value by  a factor of  3 or more, absorptive NP in the form of $(\bar s b)(\bar \tau \tau)$ operators obviously does not provide a satisfactory description of the large dimuon charge asymmetry observed by the D\O \ collaboration. This is a model-independent conclusion that can be shown to hold in explicit models of NP with modification of the $b \to s \tau^+ \tau^-$ channel such as leptoquark scenarios or $Z^\prime$ models.\hspace{1mm}\cite{Bobeth:2011st}

\section{Interplay between high- and low-$\bm{p_T}$ measurements}

The discovery of the Higgs boson with mass of around $125 \, {\rm GeV}$\hspace{1mm}\cite{ATLAS:2012gk,CMS:2012gu} combined with the direct limits on new particles rule out most of the   simple  and natural implementations of NP. At this stage, it is hence very important to examine the plethora of available experimental data and to look for hints that can guide us towards special regions where NP may still hide. Indirect probes of NP as provided by rare $B$-meson decays can play a key role in this endeavor, as will be described in the following for the case of the minimal supersymmetric SM (MSSM).

In the decoupling limit of the MSSM, {\it i.e.}, $M_A^2 \gg M_Z^2$, the lightest CP-even Higgs boson acquires the loop-corrected squared mass\hspace{1mm}\cite{Djouadi:2005gj}
\beq \label{eq:mh2}
M_h^2 \approx M_Z^2 \, c^2_{2\beta} + \frac{3  G_F}{\sqrt{2} \pi^2}\, m_t^4 \left [ - \ln \left ( \frac{m_t^2}{m_{\tilde t}^2 } \right ) + \frac{X_t^2}{m_{\tilde t}^2} \left ( 1 - \frac{X_t^2}{12 m_{\tilde t}^2} \right )  \right ] \,.
\eeq
Here $c_{2 \beta} = \cos \left ( 2 \beta \right)$,  $m_{\tilde t}^2 = m_{\tilde t_1} m_{\tilde t_2}$, and $X_t = A_t - \mu/t_\beta$  denotes the stop-mixing parameter, which  depends on the trilinear stop-Higgs boson coupling $A_t$ and the higgsino mass parameter $\mu$. From the above expression, one concludes that to raise $M_h$ from the $Z$-boson mass $M_Z$ to $125 \, {\rm GeV}$ requires sizable values of $t_\beta = \tan \beta$,  a heavy stop spectrum ($m_{\tilde t} \gtrsim 1 \, {\rm TeV}$), and/or large stop mixing ($|A_t| \gtrsim 2 \, {\rm TeV}$). 

The mentioned MSSM parameters also play an important role in the production and the decay of the Higgs. Applying Higgs low-energy theorems, it is easy to derive that the modification of the $gg \to h$ production cross section due to stops, can be written as\hspace{1mm}\cite{Dermisek:2007fi}
\beq \label{eq:Rh}
    R_h  = \frac{\sigma(gg\to h)_{\rm MSSM}}{\sigma(gg\to h)_{\rm SM}} \approx \left [ 1 +  \frac{m_t^2}{4} \left ( \frac{1}{m_{\tilde t_1}^2} + \frac{1}{m_{\tilde t_2}^2} - \frac{X_t^2}{m_{\tilde t_1}^2 m_{\tilde t_2}^2} \right )  \right ]^2 \,.
\eeq
One infers that the amount of mixing in the stop sector determines whether~(\ref{eq:Rh}) is smaller or larger than 1. For no mixing ($X_t = 0$), one has $R_h > 1$, while if $X_t$ is parametrically larger than the mass eigenvalues $m_{\tilde t_{1,2}}$  (with  $m_{\tilde t_1} < m_{\tilde t_2}$) then one has $R_h < 1$. The fact that in the MSSM, to make the Higgs sufficiently heavy, one needs large/maximal mixing, then implies that for a random MSSM parameter point with  $M_h \approx 125 \, {\rm GeV}$ one should find a suppression of $\sigma (gg \to h)$. In fact, this is precisely what happens in large parts of the parameter space. 

For $M_A \gg M_Z$ and $t_\beta \gg 1$, charged Higgs as well as chargino effects are strongly suppressed in $h \to \gamma \gamma$, but stau loops can have a notable impact on the diphoton signal strength 
\beq \label{eq:Rgamma}
R_\gamma = \frac{\big[ \sigma(pp\to h) \hspace{0.25mm} {\rm BR} (h \to \gamma \gamma )\big ]_{\rm MSSM}}{\big [ \sigma(pp\to h) \hspace{0.25mm}  {\rm BR} (h \to \gamma \gamma )\big ]_{\rm SM}}\approx  1 + 0.10 \, \frac{m_\tau^2 X_\tau^2}{m_{\tilde \tau_1}^2 m_{\tilde \tau_2}^2} \,, 
\eeq
if mixing in the stau sector is large, {\it i.e.}, $|X_\tau| = |A_\tau - \mu \hspace{0.5mm} t_\beta| \gg m_{\tilde \tau_{1,2}}$ and the stau mass eigenstate $\tilde \tau_1$ is sufficiently light.\hspace{1mm}\cite{Carena:2011aa} In (\ref{eq:Rgamma}) we have assumed that $M_h \approx 125 \, {\rm GeV}$ and explicitly included only stau effects. 

\begin{figure}[!t]
\begin{center}
\includegraphics[width=0.32 \textwidth]{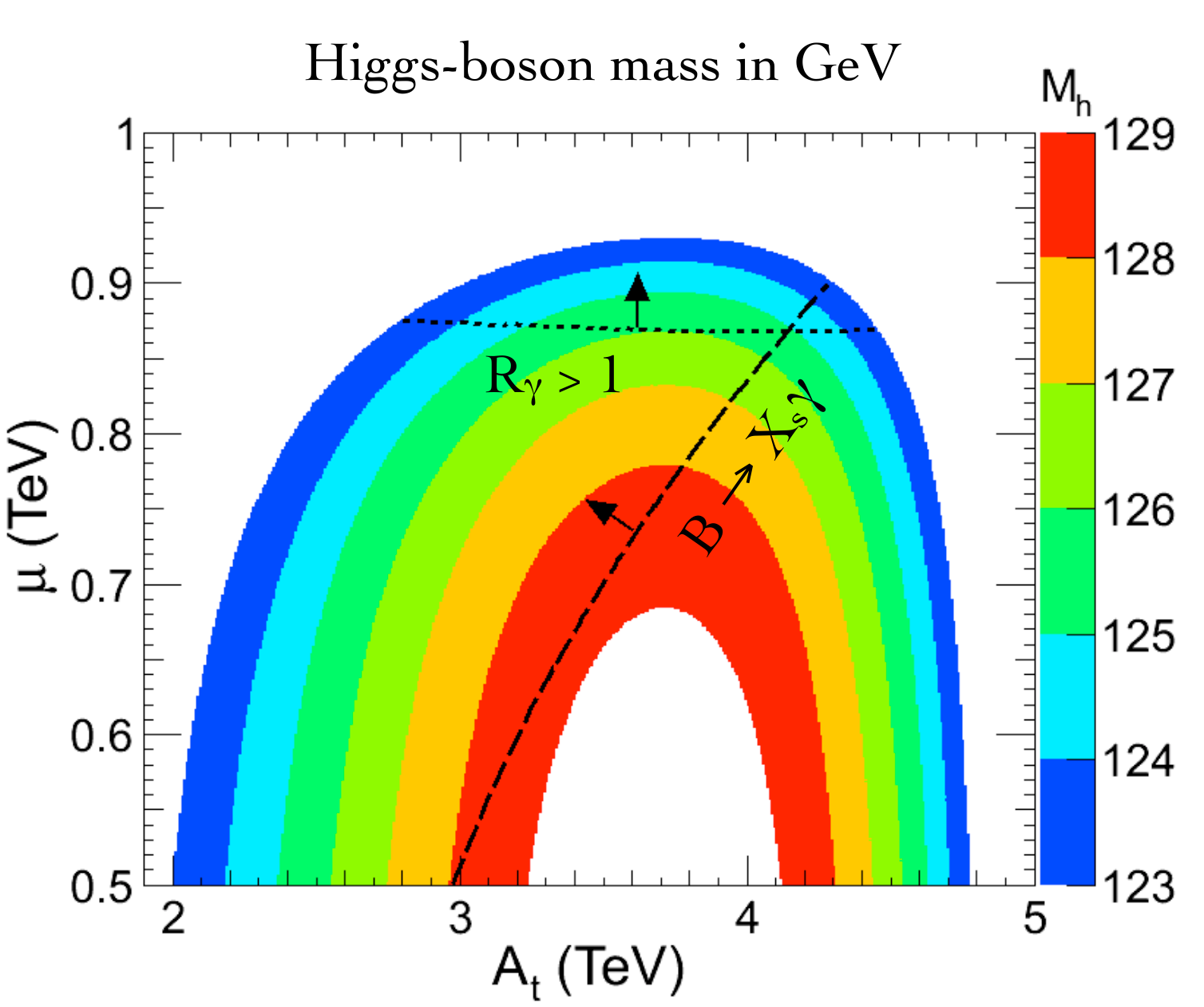} \,
\includegraphics[width=0.32 \textwidth]{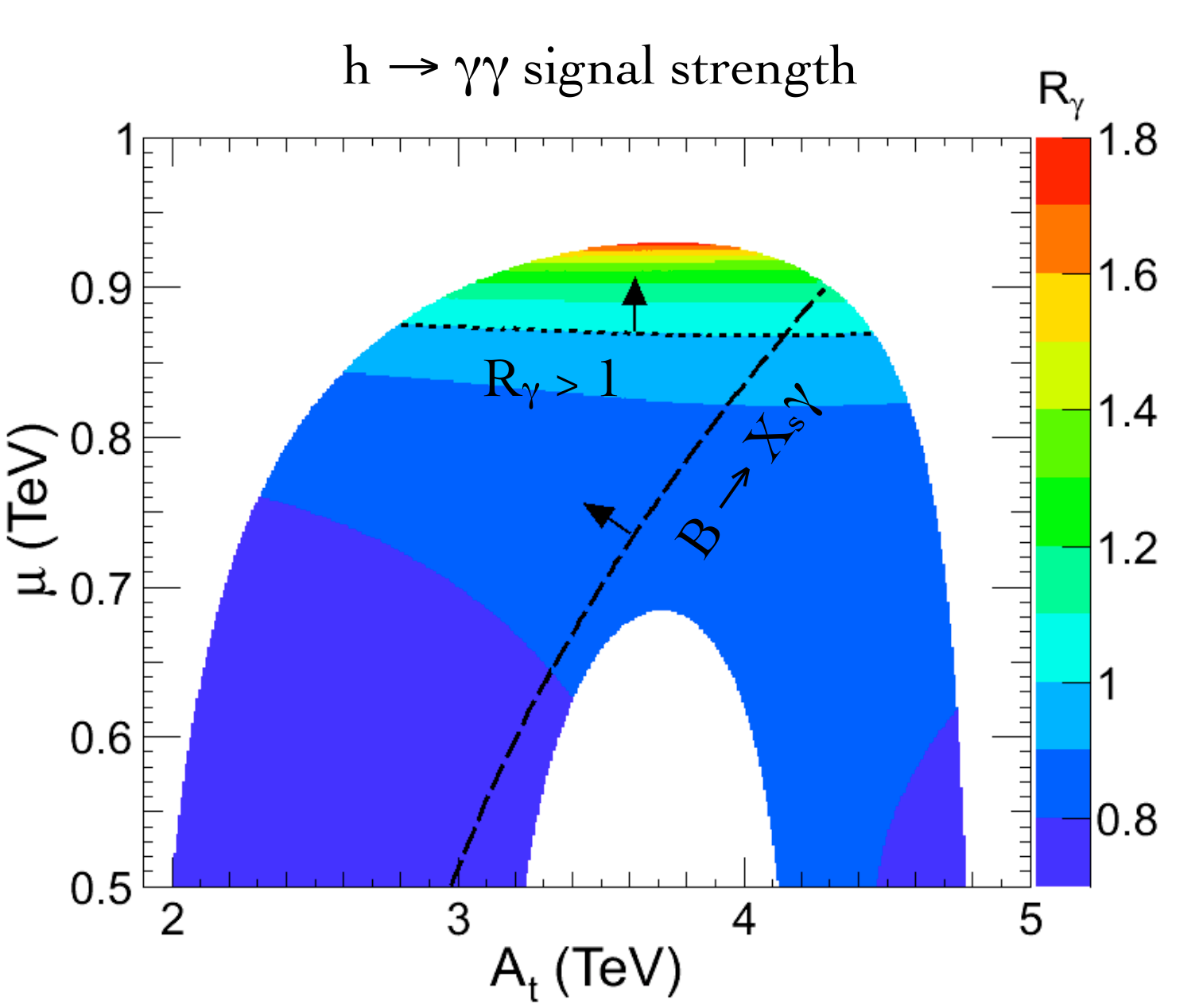} \,
\includegraphics[width=0.32 \textwidth]{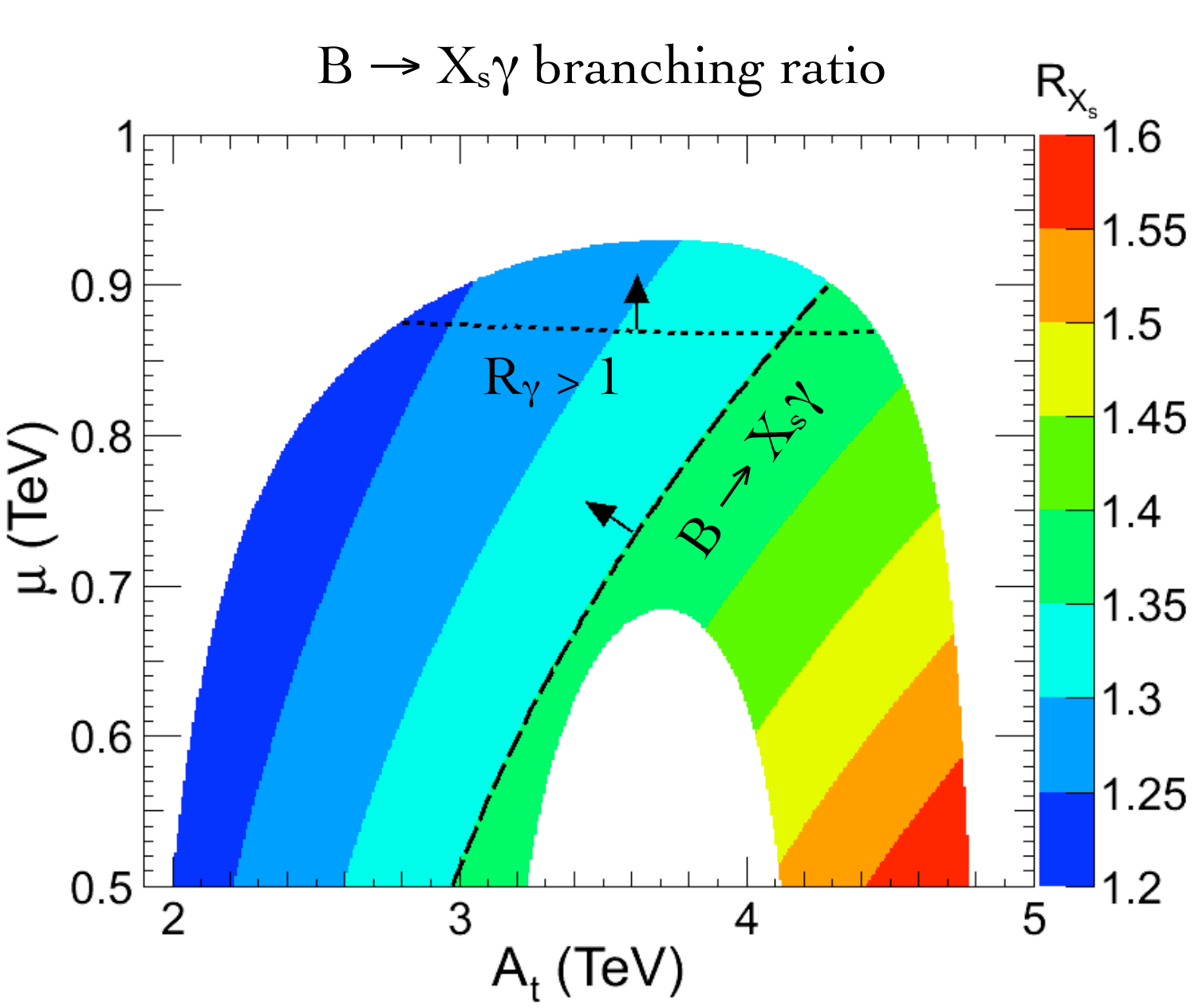} 

\includegraphics[width=0.32 \textwidth]{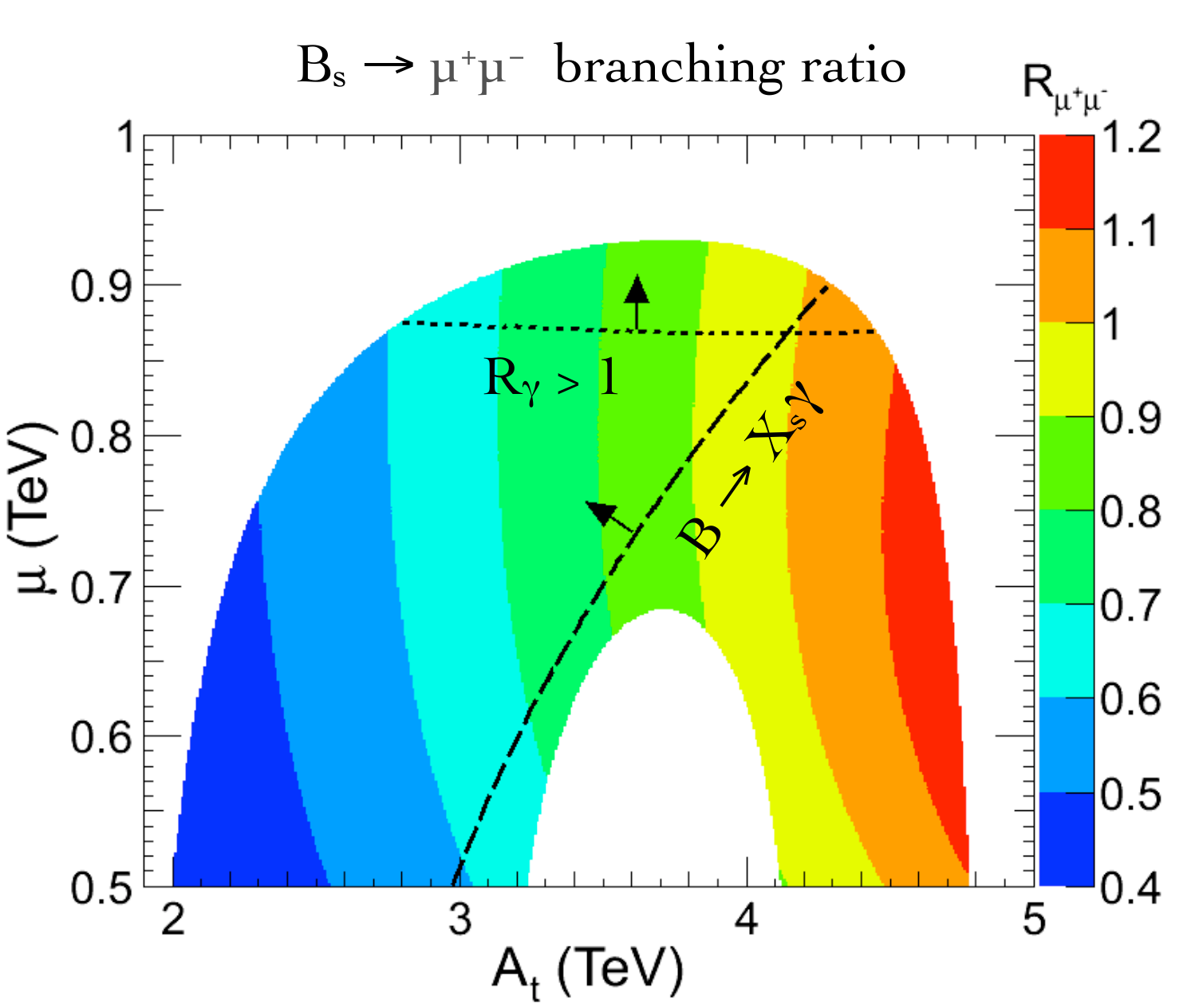} \,
\includegraphics[width=0.32 \textwidth]{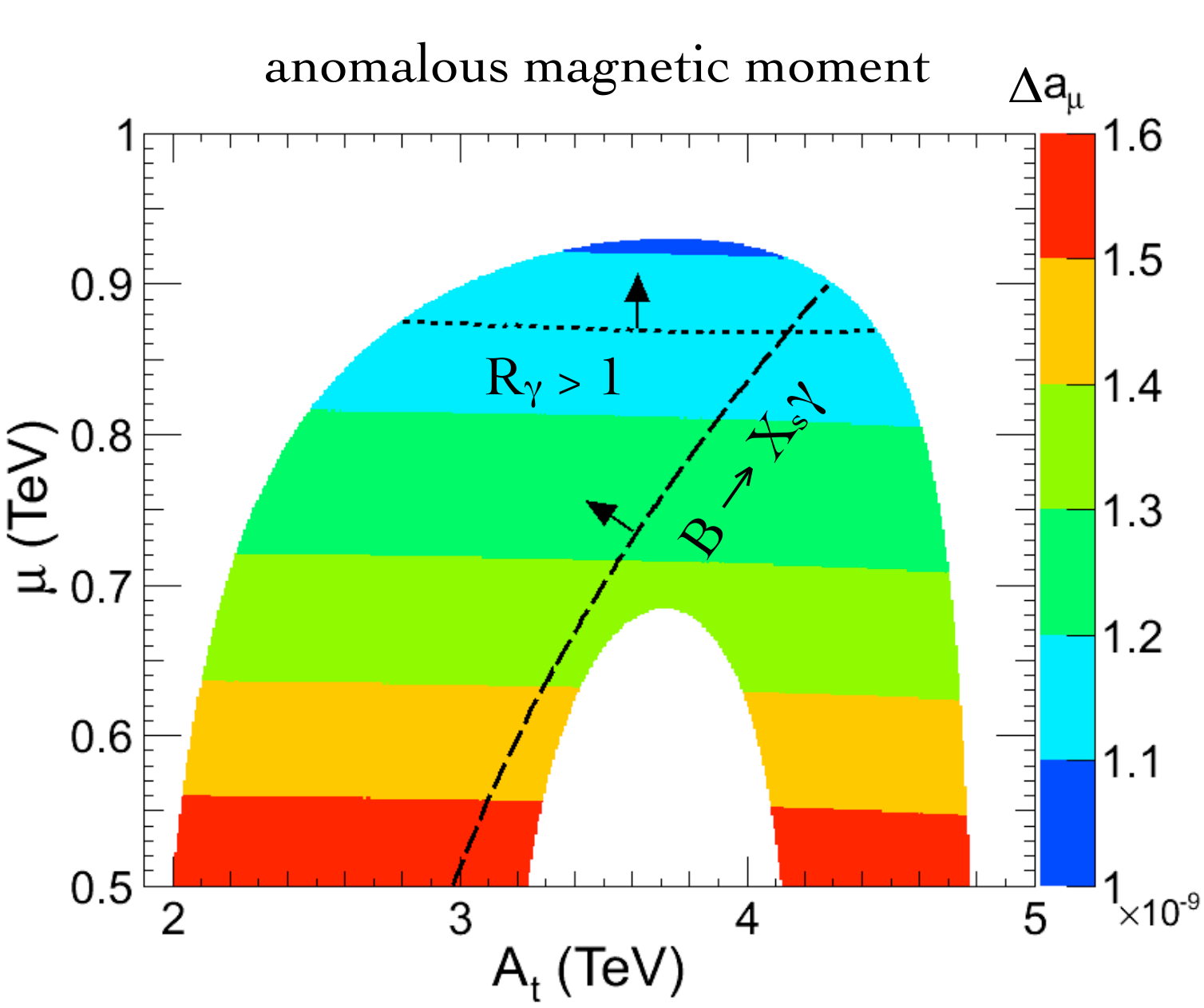} \, 
\includegraphics[width=0.32 \textwidth]{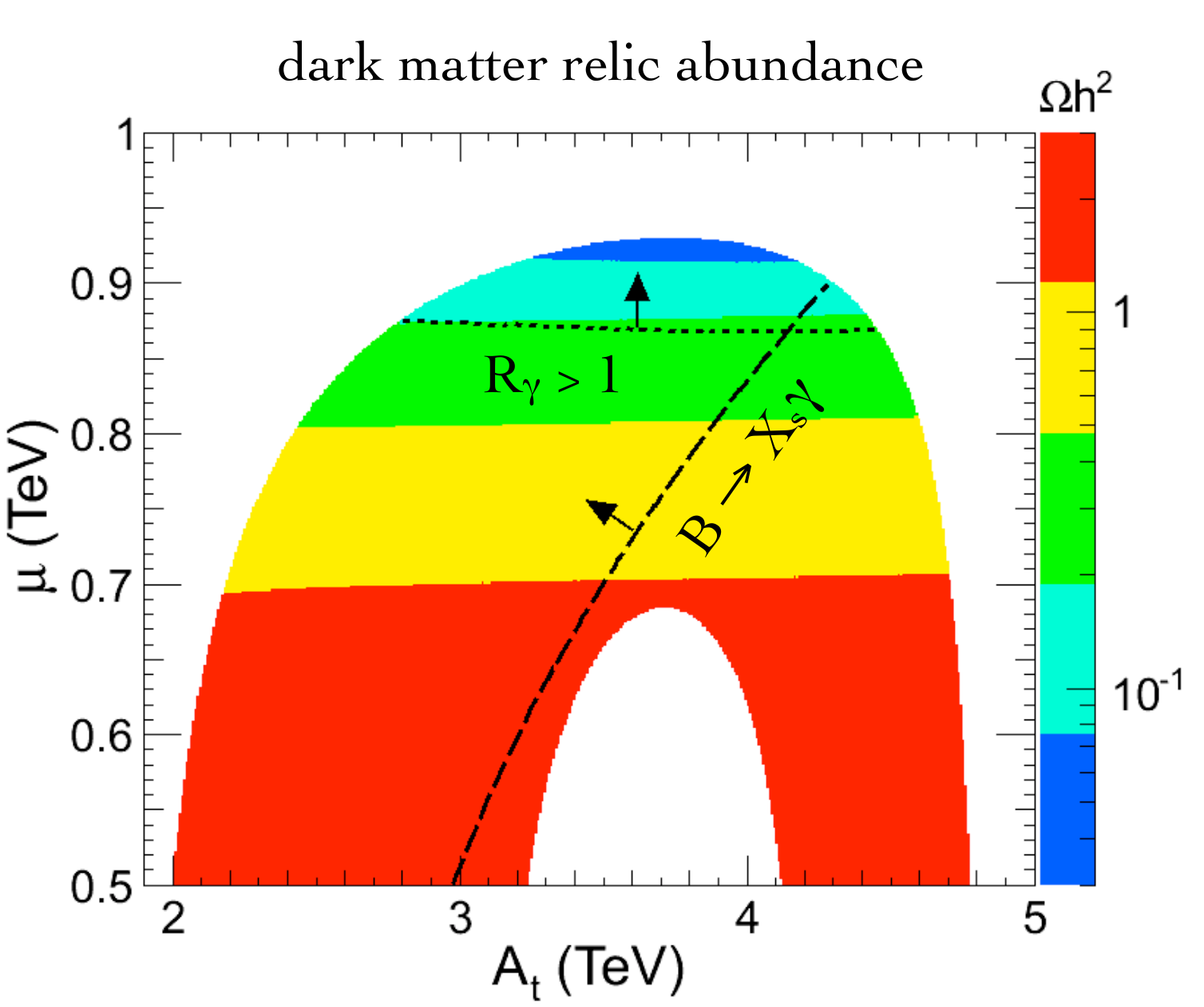}
\end{center}
\begin{center}
\caption{Predictions for $M_h$ in $\rm GeV$ (upper left), $R_{\gamma}$ (upper center), $R_{X_s}$ (upper right), $R_{\mu^+ \mu^-}$ (lower left),  $\Delta a_\mu$ (lower center), and $\Omega h^2$~(lower right). The dotted black lines indicate the parameter regions with $R_\gamma > 1$, while the dashed black lines correspond to the $95\% \,{\rm CL}$ regions favored by $B \to X_s \gamma$. See text for further explanations. \label{fig:fig1}}
\end{center}
\end{figure}  

The above discussion should have made clear that the part of the MSSM parameter space with $M_A, t_\beta \to \infty$ represents a phenomenologically interesting region for Higgs physics. How do the  dominant MSSM corrections to $B \to X_s \gamma$ and $B_s \to \mu^+ \mu^-$ 
look like in this  limiting case? For $B \to X_s \gamma$, one obtains\hspace{1mm}\cite{Haisch:2012re} 
\beq \label{eq:RXs}
R_{X_s}  =  \frac{{\rm BR} (B \to X_s \gamma)_{\rm MSSM}}{{\rm BR} (B \to X_s \gamma)_{\rm SM}}  \approx 1- 2.61 \,  \Delta C_7 + 1.66 \, ( \Delta C_7 )^2 \,,
\eeq 
where the dominant one-loop contributions are provided by diagrams with top squarks and higgsino-like charginos, leading to $\Delta C_7 \propto - \mu A_t \, t_\beta \, m_t^2/m_{\tilde t}^4$. The sign of $\Delta C_7$ combined with that of  (\ref{eq:RXs})  implies that for $\mu A_t>0$ ($\mu A_t<0$)  the MSSM branching ratio is larger (smaller) than the SM expectation. For the choices $t_\beta = 50$, $m_{\tilde t} =  1.5 \, \TeV$, $|\mu|  = 1 \, \TeV$, and $|A_t| = 3 \, {\rm TeV}$, one finds numerically an enhancement/suppression of   ${\rm BR} (B \to X_s \gamma)_{\rm MSSM}$ by ${\cal O} (30\%)$. Shifts of this size are detectable given  the present theoretical calculations and experimental extractions. 

In the case of the purely leptonic $B_s$ decay, one can write\hspace{1mm}\cite{Haisch:2012re}  
\beq \label{eq:Rmumu}
R_{\mu^+\mu^-}  =  \frac{{\rm BR} (B_s \to \mu^+ \mu^-)_{\rm MSSM}}{{\rm BR} (B_s \to \mu^+ \mu^-)_{\rm SM}} \approx  1 - 13.2  \, C_P + 43.6 \left (C_S^2 + C_P^2 \right )  \,,
\eeq
where $C_{P,S}$ denote the dimensionless Wilson coefficients of the semileptonic  pseudo-scalar and scalar operators. The term linear in $C_P$ arises from the interference with the SM contribution to the semileptonic axial-vector operator.  For $C_P >0$ it interferes destructively with the  term proportional to $\left (C_S^2 + C_P^2 \right )$, which implies that a pseudo-scalar contribution of the correct sign and size will lead to a suppression of ${\rm BR} (B_s \to \mu^+ \mu^-)_{\rm MSSM}$ below its SM value. 

Within the MSSM the contributions to  $C_{P,S}$ with the strongest $t_\beta$ dependence arise from neutral Higgs double penguins.\hspace{1mm}\cite{Babu:1999hn}  In the decoupling limit, one has  
\beq \label{eq:CP}
C_P \approx - C_S \, \propto \, \mu A_t \, \frac{t_\beta^3}{(1+\epsilon_b \hspace{0.25mm} t_\beta)^2} \; \frac{m_t^2}{m_{\tilde t}^2} \, \frac{m_b  m_\mu}{M_W^2 M_A^2} \,, \qquad \epsilon_b \propto  \frac{\alpha_s}{\pi} \frac{\mu M_3}{m_{\tilde b}^2} \,.
\eeq
Here $\epsilon_b$ encodes loop-induced non-holomorphic terms due to gluino exchange which introduce a dependence of $C_{P,S}$ on ${\rm sgn} \left (\mu M_3 \right)$. From (\ref{eq:CP}) one deduces that the sign of $C_S$ ($C_P$) is opposite to (follows) that of  $\mu A_t$ and that both coefficients are suppressed (enhanced) for $\mu M_3 > 0$ ($\mu M_3 < 0$). The recent LHCb measurement\hspace{1mm}\cite{Aaij:2012nna}  of ${\rm BR} (B_s \to \mu^+ \mu^-) = (3.2^{+1.5}_{-1.2}) \cdot 10^{-9}$ hence clearly favors $\mu A_t >0$ and  $\mu M_3 >0$, if  Higgs exchange gives the dominant contribution to~(\ref{eq:Rmumu}). 

In Figure~\ref{fig:fig1} we show the results of numerical scans in the $A_t$--$\mu$ plane.\hspace{1mm}\cite{Haisch:2012re}   We restrict ourselves to the quadrant  with $A_t >0$ and $\mu > 0$  since it shows the most interesting effects and correlations. The  predictions correspond to  $t_\beta = 60$, $M_A = 1\, {\rm TeV}$, $M_1 = 50 \, {\rm GeV}$, $M_2 = 300 \, {\rm GeV}$, $M_3 = 1.2 \, {\rm TeV}$, $\tilde m_{Q_3} = \tilde m_{u_3} = 1.5 \, {\rm TeV}$, $\tilde m_{L_3} = \tilde m_{l_3} = 350 \, {\rm GeV}$, while we take common soft  masses of $1.5 \, {\rm TeV}$ and $2 \,{\rm TeV}$ ($1 \, {\rm TeV}$) for the other  ``left-" and ``right-handed" squark (sleptons). We furthermore employ  $A_b = 2.5 \, {\rm TeV}$ and $A_\tau = 500 \, {\rm GeV}$, while the first and second generation trilinear couplings take the same values as those of the third generation. We see that for the above choice of parameters, $A_t$ has to lie in the range of $[2,5] \, {\rm TeV}$ to accommodate $M_h \in [123,129] \, {\rm GeV}$ and that $R_\gamma > 1$ can only be obtained in a narrow sliver  around $ \mu=900 \, {\rm GeV}$. For our choice of soft  masses $\tilde m_{L_3}$ and $\tilde m_{l_3}$ such large $\mu$ values lead to stau masses $m_{\tilde \tau_1} \in [80, 120] \, {\rm GeV}$. This strong correlation between notable enhancements in $R_\gamma$ and the presence of a very light stau is a smoking gun signal of the discussed scenario. From the panel showing $R_{X_s}$, one infers that ${\rm BR} (B \to X_s \gamma)$ is always enhanced (by about 20\% to 60\%) with respect to the SM. This is an interesting and potentially important finding, since the $B \to X_s \gamma$  constraint starts cutting into the already narrow region in the $A_t$--$\mu$ plane with $M_h \approx 125 \,{\rm GeV}$ and $R_\gamma > 1$. We also observe that   solutions that satisfy $B \to X_s \gamma$ typically feature a suppression of $B_s \to \mu^+ \mu^-$. In fact, asking for an agreement with $R_{X_s}$ at the $95\% \, {\rm CL}$ as well as $R_\gamma > 1$, implies $R_{\mu^+ \mu^-} \in [0.6, 1.0]$. The figure furthermore shows that assuming a light slepton spectrum, the long-standing discrepancy of the anomalous magnetic moment of the muon $a_\mu$ is reduced in  the parameter region selected by $M_h$ and the enhanced diphoton signal. Finally, notice that in this very region of parameter space also the correct  thermal dark matter relic density $\Omega h^2$ can be achieved, but only if one assumes the hierarchy $|M_1| \ll |M_2| \ll |\mu|$.  A typical MSSM spectrum leading to a significantly enhanced $h \to \gamma \gamma$ rate as well as the correct value of $\Omega h^2$, hence contains a light bino as the dark matter candidate, a light and maximally mixed stau,  and a heavy higgsino. The aforementioned  deviations and found correlations could be tested in the near future and hence may become very valuable as guidelines and consistency checks, in particular if the high-$p_T$ LHC experiments start to see supersymmetric partners.

\section{Conclusions}

While the 7 TeV and 8 TeV LHC runs took the hope to find spectacular effects in heavy flavor physics,  the low-$p_T$ data is at present not precise enough to exclude NP contaminations of ${\cal O} (50\%)$. This still leaves room for visible and interesting effects in rare $B$-meson decays, given the theoretical cleanness of these observables.  Since there is no direct sign of NP at the LHC  and  also the Higgs signal strengths look very much SM-like, such indirect NP probes  are more important than ever and only the synergy between high- and low-$p_T$ observations may give us the key to solving the puzzles of fundamental physics. The expected LHC precision measurements of the $B$-mixing observables, $B_s \to \mu^+ \mu^-$, $B \to K^\ast \ell^+ \ell^-$, the angle $\gamma$, {\it etc.} may play a crucial  role in this endeavor.

\section*{Acknowledgments}

I am grateful to the organizers of the BSM conference Vietnam 2012 for the invitation to a great conference, and to Marc~Besancon  for encouraging me to write these proceedings. Travel support from the UNILHC network (PITN-GA-2009-237920) is acknowledged.

\end{document}